\documentclass[aps,preprint,nofootinbib,floatfix]{revtex4}%
\usepackage{amssymb}
\usepackage{amsmath}
\usepackage{epsfig}
\usepackage{amsfonts}
\usepackage{graphicx}%
\begin{document}
\title{Power-law Behavior of Strings Scattered from Domain-wall at High Energies and
Breakdown of their Linear Relations }
\author{Chuan-Tsung Chan}
\email{ctchan@thu.edu.tw}
\affiliation{Department of Physics, Tung-Hai University, Taichung, Taiwan, R.O.C.}
\author{Jen-Chi Lee}
\email{jcclee@cc.nctu.edu.tw}
\affiliation{Department of Electrophysics, National Chiao-Tung University, Hsinchu, Taiwan, R.O.C.}
\author{Yi Yang}
\email{yyang@phys.cts.nthu.edu.tw}
\affiliation{Department of Electrophysics, National Chiao-Tung University and Physics
Division, National Center for Theoretical Sciences, Hsinchu, Taiwan, R.O.C.}
\date{\today }

\begin{abstract}
In contrast to the common wisdom, we discover that, instead of the exponential
fall-off of the form factors with Regge-pole structure, the high energy
scattering amplitudes of string scattered from Domain-wall behave as power-law
with Regge-pole structure. This is to be compared with the well-known
power-law form factors without Regge-pole structure of the D-instanton
scatterings. This discovery makes Domain-wall scatterings an unique example of
a hybrid of string and field theory scatterings. The calculation is done for
bosonic string scatterings of arbitrary massive string states from D24-brane.
Moreover, we discover that the usual linear relations of high-energy string
scattering amplitudes at each fixed mass level break down for the Domain-wall
scatterings. This result gives a strong evidence that the existence of the
infinite linear relations, or stringy symmetries, of high-energy string
scattering amplitudes is responsible for the softer, exponential fall-off high
energy string scatterings than the power-law field theory scatterings.

\end{abstract}
\maketitle

\address{Department of Electrophysics, National Chiao Tung University,
Hsinchu 300, Taiwan}

%\draft
%\wideabs{

In contrast to the local quantum field theory, string theory is remarkable for
its ultraviolet finiteness. Being a consistent quantum theory of gravity with
no free parameter, it is conceivable that an huge symmetry gets restored at
high energies. Recently high energy, fixed angle behavior of string scattering
amplitudes \cite{GM, Gross, GrossManes} was intensively reinvestigated for
massive string states at arbitrary mass levels \cite{ChanLee1,ChanLee2,
CHL,CHLTY,PRL,paperB,susy,Closed}. The motivation was to uncover the
fundamental hidden stringy spacetime symmetry conjectured by Gross in 1988. An
important new ingredient of this approach is the zero-norm states (ZNS)
\cite{ZNS1,ZNS3,ZNS2} in the old covariant first quantized (OCFQ) string
spectrum. One utilizes the decoupling of ZNS in the high energy limit to
obtain infinite linear relations, or stringy symmetries, among string
scattering amplitudes. Moreover, these linear relations can be used to fix the
ratios among high energy scattering amplitudes of different string states at
each fixed mass level algebraically. This explicitly shows that there is only
one independent component of high-energy scattering amplitude at each mass
level. On the other hand, a saddle-point method was developed to calculate the
general formula of tree-level high-energy scattering amplitudes of four
arbitrary string states to verify the ratios calculated above. This general
formula expresses all high-energy string scattering amplitudes in terms of
four tachyon scattering amplitude as conjectured by Gross \cite{Gross}.

In addition to the infinite linear relations, there are two other fundamental
characteristics of high energy string scattering amplitudes, which make them
very different from those of field theory scatterings. The first one is the
exponential fall-off behavior of the form factors of string scatterings in the
high-energy limit in contrast to the power-law behavior of point-particle
field theory scatterings. It is the main result of this letter that the
infinite linear relations, or stringy symmetries, among high-energy string
scattering amplitudes of different string states discussed above are
responsible for this softer exponential fall-off behavior of string
scatterings than the power-law behavior of field theory scatterings. The
second one is the Regge-pole structure \cite{Closed} in the high energy string
scattering amplitudes due to the infinite number of resonances in the string
spectrum. In the scattering processes of string scattered from D-brane
\cite{Klebanov,Myers}, it was claimed that the exponential fall-off form
factors with Regge-pole structure is a general feature of all D$p$-brane
scatterings except D-instanton, which exhibits power-law form factor without
Regge-pole structure, and thus resembles a field theory amplitude instead.

In our recent work \cite{Dscatt}, we have shown that the linear relations
stated above persist for the string/D$p$-brane scatterings with $p\geqslant0.$
The linear relations for the D-particle scatterings were explicitly
demonstrated. All the high energy string/D$p$-brane scattering amplitudes with
$p\geqslant0$ behave as exponential fall-off as was claimed in Ref
\cite{Klebanov}. In this letter, in contrast to the common wisdom, we discover
that, instead of the exponential fall-off behavior of the form factors with
Regge-pole structure, the high-energy scattering amplitudes of string
scattered from D$24$-brane, or Domain-wall, behave as \textit{power-law} with
Regge-pole structure. This is to be compared with the well-known power-law
form factors without Regge-pole structure of the D-instanton scatterings
stated above. This discovery makes Domain-wall scatterings an unique example
of a hybrid of string and field theory scatterings. Our calculation will be
done for bosonic string scatterings of arbitrary massive string states from
D$24$-brane. Moreover, we discover that the usual linear relations
\cite{Dscatt} of high energy string scattering amplitudes at each fixed mass
level breaks down for the Domain-wall scatterings. This result gives a strong
evidence that the existence of the infinite linear relations, or stringy
symmetries, of high energy string scattering amplitudes is responsible for the
softer, exponential fall-off high-energy string scatterings than the power-law
field theory scatterings.

At a fixed mass level $M_{op}^{2}=2(n-1)$ of 26D open bosonic string theory,
it was shown that \cite{CHLTY,PRL} a four-point function is at the leading
order at high-energy limit only for states of the following form ( we use the
notation of \cite{GSW})%
\begin{equation}
\left\vert n,2m,q\right\rangle \equiv(\alpha_{-1}^{T})^{n-2m-2q}(\alpha
_{-1}^{L})^{2m}(\alpha_{-2}^{L})^{q}\left\vert 0,k\right\rangle . \label{1}%
\end{equation}
where $n\geqslant2m+2q,m,q\geqslant0.$Note that, in the high energy limit, the
scattering process becomes a plane scattering (however, see the Domain-wall
scattering in the end of this letter). The state in Eq.(\ref{1}) is
arbitrarily chosen to be the second vertex of the four-point function. The
other three points can be any string states. We have defined the normalized
polarization vectors of the second string state to be \cite{ChanLee1,ChanLee2}%
\begin{equation}
e_{P}=\frac{1}{M_{op}}(E_{2},\mathrm{k}_{2},0)=\frac{k_{2}}{M_{op}}, \label{2}%
\end{equation}%
\begin{equation}
e_{L}=\frac{1}{M_{op}}(\mathrm{k}_{2},E_{2},0), \label{3}%
\end{equation}%
\begin{equation}
e_{T}=(0,0,1) \label{4}%
\end{equation}
in the CM frame contained in the plane of scattering. By using the decoupling
of two types of ZNS in the high energy limit, it was shown that there exists
infinite linear relations among string scattering amplitudes \cite{CHLTY,PRL}%
\begin{equation}
\mathcal{T}^{(n,2m,q)}=\left(  -\frac{1}{M_{op}}\right)  ^{2m+q}\left(
\frac{1}{2}\right)  ^{m+q}(2m-1)!!\mathcal{T}^{(n,0,0)}. \label{5}%
\end{equation}
Moreover, these linear relations can be used to fix the ratios among high
energy scattering amplitudes of different string states at each fixed mass
level algebraically. Eq.(\ref{5}) explicitly shows that there is only one
independent high-energy scattering amplitudes at each fixed mass level.

We now turn to the case of D-brane scatterings. In our recent work
\cite{Dscatt}, we discover that scatterings of bosonic massive closed string
states at arbitrary mass levels from D-brane can be expressed in terms of the
generalized hypergeometric function $_{3}F_{2}$ with special arguments, which
terminates to a finite sum and, as a result, the whole scattering amplitudes
consistently reduce to the usual beta function. For the simple case of
D-particle, we explicitly calculated high-energy scattering amplitude of an
incoming closed string tensor state $\left(  \alpha_{-1}^{T}\right)
^{n-2q}\left(  \alpha_{-2}^{L}\right)  ^{q}\otimes\left(  \tilde{\alpha}%
_{-1}^{T}\right)  ^{n-2q^{\prime}}\left(  \tilde{\alpha}_{-2}^{L}\right)
^{q^{\prime}}\left\vert 0\right\rangle $ with $M^{2}\equiv\frac{M_{closed}%
^{2}}{2\alpha_{closed}^{\prime}}=2(n-1)=-k_{2}^{2}$ and an outgoing tachyon
state to be \cite{Dscatt}%
\begin{align}
&  A_{D-Par}\nonumber\\
&  =A_{D-Par}^{\left(  0\rightarrow1\right)  }+A_{D-Par}^{\left(
1\rightarrow\infty\right)  }\nonumber\\
&  =2\left(  -1\right)  ^{a_{0}}E^{2n}\left(  -\dfrac{1}{2M}\right)
^{q+q^{\prime}}B\left(  a_{0}+1,\frac{b_{0}+1}{2}\right) \nonumber\\
&  =2\left(  -1\right)  ^{a_{0}}E^{2n}\left(  -\dfrac{1}{2M}\right)
^{q+q^{\prime}}\frac{\Gamma(a_{0}+1)\Gamma(\frac{b_{0}+1}{2})}{\Gamma
(a_{0}+\frac{b_{0}}{2}+\frac{3}{2})}, \label{8}%
\end{align}
where the high energy beta function $B(a_{0}+1,\frac{b_{0}+1}{2})$ is
independent of $q+q^{\prime}$, and $a_{0}$, $b_{0}$ will be defined in
Eqs.(\ref{22}) and (\ref{23}). Since the calculation of decoupling of high
energy ZNS without D-brane remains the same as the calculation of string
scattered from D-brane, it is a remarkable result that the ratios $\left(
-\dfrac{1}{2M}\right)  ^{q+q^{\prime}}$in Eq.(\ref{8}) for different high
energy scattering amplitudes at each fixed mass level is consistent with
Eq.(\ref{5}) for the scattering without D-brane as we found before
\cite{CHLTY}. One notes that the exponential fall-off behavior in energy $E$
is hidden in the high energy beta function. Since the arguments of
$\Gamma(a_{0}+1)$ and $\Gamma(a_{0}+\frac{b_{0}}{2}+\frac{3}{2})$ in
Eq.(\ref{8}) are negative in the high-energy limit, one needs to use the well
known formula%
\begin{equation}
\Gamma\left(  x\right)  =\frac{\pi}{\sin\left(  \pi x\right)  \Gamma\left(
1-x\right)  } \label{gama}%
\end{equation}
to calculate the large negative $x$ expansion of these $\Gamma$ functions, and
obtain the Regge-pole structure \cite{Closed} of the amplitude.

Presumably, other similar linear relations as Eq.(\ref{8}) for high energy
scatterings can be derived for D$p$-brane with $p\geq$ $1$. However, in the
following, in contrast to the common wisdom \cite{Klebanov}, we will show
that, instead of the exponential fall-off of the form factors with Regge-pole
structure, the high-energy scattering amplitudes of string scattered from
D$24$-brane, i.e. Domain-wall, behave like \textit{power-law} with Regge-pole
structure. Moreover, we discover that the expected linear relations of
high-energy string scattering amplitudes at each fixed mass level,
Eq.(\ref{5}) or Eq.(\ref{8}), break down for the Domain-wall scatterings. We
consider an incoming tachyon closed string state with momentum $k_{1}$and an
angle of incidence $\phi$ and an outgoing massive closed string state $\left(
\alpha_{-1}^{T}\right)  ^{n-2q}\left(  \alpha_{-2}^{L}\right)  ^{q}%
\otimes\left(  \tilde{\alpha}_{-1}^{T}\right)  ^{n-2q^{\prime}}\left(
\tilde{\alpha}_{-2}^{L}\right)  ^{q^{\prime}}\left\vert 0\right\rangle $ with
momentum $k_{2}$ and an angle of reflection $\theta$. The kinematic setup is%
\begin{align}
e^{P}  &  =\frac{1}{M}\left(  -E,\mathrm{k}_{2}\cos\theta,-\mathrm{k}_{2}%
\sin\theta\right)  =\frac{k_{2}}{M},\label{10}\\
e^{L}  &  =\frac{1}{M}\left(  -\mathrm{k}_{2},E\cos\theta,-E\sin\theta\right)
,\label{11}\\
e^{T}  &  =\left(  0,\sin\theta,\cos\theta\right)  ,\label{12}\\
k_{1}  &  =\left(  E,-\mathrm{k}_{1}\cos\phi,-\mathrm{k}_{1}\sin\phi\right)
,\label{13}\\
k_{2}  &  =\left(  -E,\mathrm{k}_{2}\cos\theta,-\mathrm{k}_{2}\sin
\theta\right)  . \label{14}%
\end{align}
In the high energy limit, the angle of incidence $\phi$ is identified to the
angle of reflection $\theta$, and $e^{P}$ approaches $e^{L}$, $\mathrm{k}%
_{1},\mathrm{k}_{2}\simeq E$. For the case of Domain-wall scattering $Diag$
$D_{\mu\nu}=(-1,1,-1)$, and we have%
\begin{align}
a_{0}  &  \equiv k_{1}\cdot D\cdot k_{1}\sim-2E^{2}\sin^{2}\phi-2M_{1}^{2}%
\cos^{2}\phi+M_{1}^{2},\label{22}\\
b_{0}  &  \equiv2k_{1}\cdot k_{2}+1\nonumber\\
&  \sim4E^{2}\sin^{2}\phi+4M_{1}^{2}\cos^{2}\phi-\left(  M_{1}^{2}%
+M^{2}\right)  +1, \label{23}%
\end{align}
The scattering amplitude can be calculated to be%

\begin{align}
A_{D-Wall}  &  =\left(  -1\right)  ^{q+q^{\prime}}\int d^{2}z_{1}d^{2}%
z_{2}\left(  z_{1}-\bar{z}_{1}\right)  ^{k_{1}\cdot D\cdot k_{1}}\left(
z_{2}-\bar{z}_{2}\right)  ^{k_{2}\cdot D\cdot k_{2}}\left\vert z_{1}%
-z_{2}\right\vert ^{2k_{1}\cdot k_{2}}\left\vert z_{1}-\bar{z}_{2}\right\vert
^{2k_{1}\cdot D\cdot k_{2}}\nonumber\\
&  \cdot\left[  \frac{ie^{T}\cdot k_{1}}{z_{1}-z_{2}}+\frac{ie^{T}\cdot D\cdot
k_{1}}{\bar{z}_{1}-z_{2}}+\frac{ie^{T}\cdot D\cdot k_{2}}{\bar{z}_{2}-z_{2}%
}\right]  ^{n-2q}\nonumber\\
&  \cdot\left[  \frac{ie^{T}\cdot D\cdot k_{1}}{z_{1}-\bar{z}_{2}}%
+\frac{ie^{T}\cdot k_{1}}{\bar{z}_{1}-\bar{z}_{2}}+\frac{ie^{T}\cdot D\cdot
k_{2}}{z_{2}-\bar{z}_{2}}\right]  ^{n-2q^{\prime}}\nonumber\\
&  \cdot\left[  \frac{e^{L}\cdot k_{1}}{\left(  z_{1}-z_{2}\right)  ^{2}%
}+\frac{e^{L}\cdot D\cdot k_{1}}{\left(  \bar{z}_{1}-z_{2}\right)  ^{2}}%
+\frac{e^{L}\cdot D\cdot k_{2}}{\left(  \bar{z}_{2}-z_{2}\right)  ^{2}%
}\right]  ^{q}\left[  \frac{e^{L}\cdot D\cdot k_{1}}{\left(  z_{1}-\bar{z}%
_{2}\right)  ^{2}}+\frac{e^{L}\cdot k_{1}}{\left(  \bar{z}_{1}-\bar{z}%
_{2}\right)  ^{2}}+\frac{e^{L}\cdot D\cdot k_{2}}{\left(  z_{2}-\bar{z}%
_{2}\right)  ^{2}}\right]  ^{q^{\prime}}. \label{25}%
\end{align}
Set $z_{1}=iy$ and $z_{2}=i$ to fix the $SL(2,R)$ gauge, and include the
Jacobian $d^{2}z_{1}d^{2}z_{2}\rightarrow4\left(  1-y^{2}\right)  dy$, we
have, for the $\left(  0\rightarrow1\right)  $ channel,%
\begin{align}
&  A_{D-Wall}^{(0\rightarrow1)}\nonumber\\
&  \simeq4\left(  2i\right)  ^{k_{1}\cdot D\cdot k_{1}+k_{2}\cdot D\cdot
k_{2}}\left(  \frac{E\sin2\phi}{2}\right)  ^{2n}\left(  \frac{1}{2M\cos
^{2}\phi}\right)  ^{q+q^{\prime}}\nonumber\\
&  \cdot\sum_{i=0}^{q+q^{\prime}}\binom{q+q^{\prime}}{i}2^{i}\int_{0}%
^{1}dy\text{ }y^{k_{2}\cdot D\cdot k_{2}}\left(  1-y\right)  ^{2k_{1}\cdot
k_{2}+1}\nonumber\\
&  \cdot\left(  1+y\right)  ^{2k_{1}\cdot D\cdot k_{2}+1}\left[  \frac
{1+y}{1-y}\right]  ^{2n-\left(  q+q^{\prime}\right)  }\left(  \frac{1}%
{1-y}\right)  ^{i}\nonumber\\
&  \simeq\left(  \frac{E\sin2\phi}{2}\right)  ^{2n}\left(  \frac{1}{2M\cos
^{2}\phi}\right)  ^{q+q^{\prime}}\nonumber\\
&  \cdot\sum_{i=0}^{q+q^{\prime}}\binom{q+q^{\prime}}{i}\cdot B\left(
a_{0}+1,\frac{b+1}{2}\right)  F_{i} \label{27}%
\end{align}
where%
\begin{align}
b  &  =b_{0}+n_{b}=b_{0}-2n+\left(  q+q^{\prime}\right)  -i,\label{28}\\
F_{i}  &  \equiv\left(  1+\sqrt{\left\vert \dfrac{b}{2a_{0}+b}\right\vert
}\right)  ^{i}+\left(  1-\sqrt{\left\vert \dfrac{b}{2a_{0}+b}\right\vert
}\right)  ^{i}\label{fi2}\\
\simeq &  \left[  \left(  1+2C_{i}E\sin\phi\right)  ^{i}+\left(  1-2C_{i}%
E\sin\phi\right)  ^{i}\right]  \label{fi}%
\end{align}
with%
\begin{equation}
C_{i}\equiv\sqrt{\left\vert \frac{1}{M_{1}^{2}-M^{2}+1-2n+\left(  q+q^{\prime
}\right)  -i}\right\vert }. \label{30}%
\end{equation}
$F_{i}$ in Eq.(\ref{fi2}) is the high energy limit of the generalized
hypergeometric function $_{3}F_{2}\left(  \frac{b+1}{2},-\left[  \frac{i}%
{2}\right]  ,\dfrac{1}{2}-\left[  \frac{i}{2}\right]  ;a_{0}+\frac{b+3}%
{2},\dfrac{1}{2};1\right)  $ \cite{Dscatt}. At this stage, it is crucial to
note that%
\begin{equation}
b\simeq b_{0}\simeq-2a_{0} \label{match}%
\end{equation}
in the high energy limit for the Domain-wall scatterings. As a result, $F_{i}$
reduces to the form of Eq.(\ref{fi}), and depends on the energy $E.$ Thus in
contrast to the generic D$p$-brane scatterings with $p\geq0$, which contain
two independent kinematic variables, there is only one kinematic variable for
the special case of Domain-wall scatterings. As we will see in the following
calculation, this peculiar property will reduce the high energy beta function
in Eq.(\ref{27}) from exponential to power-law behavior and, simutaneously,
breaks down the linear relations as we had in Eq.(\ref{8}) for the D-particle
scatterings. Another way to see this breakdown is by examining Eq.(\ref{5}).
One notes that the argument in the paragraph after Eq.(\ref{1}) for scattering
with two kinematic variables, which leads to Eq.(\ref{5}), is no longer valid
as there is only one kinematic variable for the Domain-wall scatterings. It
thus becomes meaningless to study high-energy, fixed angle scattering process
for the Domain-wall scatterings. Finally, the scattering amplitude for the
$(0\rightarrow1)$ channel can be calculated to be (similar result can be
obtained for the $(1\rightarrow\infty)$ channel)%
\begin{align}
&  A_{D-Wall}^{(0\rightarrow1)}\nonumber\\
&  \simeq\left(  \frac{E\sin2\phi}{2}\right)  ^{2n}\left(  \frac{1}{2M\cos
^{2}\phi}\right)  ^{q+q^{\prime}}B\left(  a_{0}+1,\frac{b_{0}+1}{2}\right)
\nonumber\\
&  \cdot\sum_{i=0}^{q+q^{\prime}}\binom{q+q^{\prime}}{i}\cdot\dfrac{\left(
a_{0}\right)  _{0}\left(  \dfrac{b_{0}}{2}\right)  _{n_{b}/2}}{\left(
a_{0}+\dfrac{b_{0}}{2}\right)  _{n_{b}/2}}\left(  1+2C_{i}E\sin\phi\right)
^{i}\label{frac}\\
&  \simeq\left(  \frac{\cos\phi}{\sqrt{2}}\right)  ^{2n}\left(  \frac
{E\sin\phi}{M\sqrt{\left\vert M_{1}^{2}-2M^{2}-1\right\vert }\cos^{2}\phi
}\right)  ^{q+q^{\prime}}\nonumber\\
&  \cdot\frac{\Gamma(a_{0}+1)\Gamma(\frac{b_{0}+1}{2})}{\Gamma(a_{0}%
+\frac{b_{0}}{2}+\frac{3}{2})}\dfrac{1}{\left(  \frac{M_{1}^{2}-M^{2}+1}%
{2}\right)  _{-n}} \label{am}%
\end{align}
where $(\alpha)_{n}\equiv\frac{\Gamma(\alpha+n)}{\Gamma(\alpha)}$ for integer
$n$. On the other hand, since the argument of $\Gamma(a_{0}+1)$ in
Eq.(\ref{am}) is negative in the high-energy limit, we have, by using
Eq.(\ref{gama}) and Eq.(\ref{match}),%
\begin{align}
\frac{\Gamma(a_{0}+1)\Gamma(\frac{b_{0}+1}{2})}{\Gamma(a_{0}+\frac{b_{0}}%
{2}+\frac{3}{2})}  &  \simeq\frac{\pi}{\sin\left(  \pi a_{0}\right)
\Gamma\left(  -a_{0}\right)  }\frac{\Gamma(\frac{b_{0}+1}{2})}{\Gamma
(\frac{M_{1}^{2}-M^{2}+1}{2})}\nonumber\\
&  \sim\frac{1}{\sin\left(  \pi a_{0}\right)  }\frac{1}{(E\sin\phi)^{2(n-1)}}.
\label{pole}%
\end{align}
Note that the $\sin\left(  \pi a_{0}\right)  $ factor in the denominator of
Eq.(\ref{pole}) gives the Regge-pole structure, and the energy dependence
$E^{-2(n-1)}$ gives the power-law behavior in the high energy limit. As a
result, the scattering amplitude for the Domain-wall in Eq.(\ref{am}) behaves
like power-law with the Regge-pole structure. The crucial differences between
the Domain-wall scatterings in Eq.(\ref{am}) and the D-particle scatterings
(or any other D$p$-brane scatterings except Domain-wall and D-instanton
scatterings) in Eq.(\ref{8}) is the kinematic relation Eq.(\ref{match}). For
the case of D-particle scatterings \cite{Dscatt}, the corresponding factors
for both $F_{i}$ in Eq.(\ref{fi2}) and the fraction in Eq.(\ref{frac}) are
independent of energy in the high energy limit, and, as a result, the
amplitudes contain no $q+q^{\prime}$ dependent energy power factor. So one
gets the high energy linear relations for the D-particle scattering
amplitudes. On the contrary, for the case of Domain-wall scatterings, both
$F_{i}$ in Eq.(\ref{fi}) and the fraction in Eq.(\ref{frac}) depend on energy
due to the condition Eq.(\ref{match}). The summation in Eq.(\ref{frac}) is
then dominated by the term $i=q+q^{\prime}$, and the whole scattering
amplitude Eq.(\ref{am}) contains a $q+q^{\prime}$ dependent energy power
factor. As a result, the usual linear relations for the high energy scattering
amplitudes break down for the Domain-wall scatterings. It is crucial to note
that the mechanism, Eq.(\ref{match}), to drive the exponential fall-off form
factor of the D-particle scatterings to the power-law one of the Domain-wall
scatterings is exactly the same as the mechanism to break down the expected
linear relations for the domain-wall scatterings in the high energy limit. In
conclusion, this result gives a strong evidence that the existence of the
infinite linear relations, or stringy symmetries, of high-energy string
scattering amplitudes is responsible for the softer, exponential fall-off
high-energy string scatterings than the power-law field theory scatterings.

Another interesting case of D-brane scatterings is the massless form factor of
scatterings of D-instanton \cite{Klebanov}%
\begin{equation}
\frac{\Gamma(s)\Gamma(t)}{\Gamma(s+t+1)}\rightarrow\frac{1}{st},\text{ as
}s\rightarrow0\text{,} \label{st}%
\end{equation}
which contains no Regge-pole structure. In Eq.(\ref{st}), $s,t$ are the
Mandelstam variables. Eq.(\ref{st}) can be easily generalized to the
scatterings of arbitrary massive string states in the high energy limit. To
compare the D-instanton scatterings with the Domain-wall scatterings in
Eq.(\ref{pole}), one notes that in both cases there is only one kinematic
variable and, as a result, behave as power-law at high energies \cite{Decay}.
On the other hand, since $t$ is large negative in the high energy limit
\cite{Closed}, the application of Eq.(\ref{gama}) to Eq.(\ref{st}) produces no
$\sin\left(  \pi a_{0}\right)  $ factor in contrast to the Domain-wall
scatterings. So there is no Regge-pole structure for the D-instanton
scatterings. We conclude that the very condition of Eq.(\ref{match}) makes
Domain-wall scatterings an unique example of a hybrid of string and field
theory scatterings.

This work is supported in part by the National Science Council, 50 billions
project of MOE and National Center for Theoretical Science, Taiwan, R.O.C.

\end{document}